\documentclass[conference]{IEEEtran}
\usepackage[T1]{fontenc}
\usepackage{lmodern}
\usepackage{epsfig}
\usepackage{rotating}
\usepackage{amsmath}
\usepackage{amsfonts}
\usepackage{amssymb}
\usepackage{graphics}
\begin{document}
\title{Achievable rates for transmission of discrete constellations over the Gaussian MAC channel}

\author{Rizwan Ghaffar\IEEEauthorrefmark{1}, Dimitris Toumpakaris\IEEEauthorrefmark{2} and \IEEEauthorblockN{Jungwon Lee\IEEEauthorrefmark{3}}
\IEEEauthorblockA{\IEEEauthorrefmark{1}Eurecom, 2229 route des Cr\^etes  B.P.193\\ 06904 Sophia Antipolis Cedex FRANCE, Email: rizwan.ghaffar@eurecom.fr}
\IEEEauthorblockA{\IEEEauthorrefmark{2}Wireless Telecommunications Laboratory, Department of Electrical and Computer Engineering \\ University of Patras, Rio, Greece 265 00, Email: dtouba@upatras.gr}
\IEEEauthorblockA{\IEEEauthorrefmark{3}Mobile Solutions Lab, Samsung US, R\&D Center \\4921 Directors Pl., San Diego, CA 92121, Email: jungwon@stanfordalumni.org}}

\maketitle
\begin{abstract}
In this paper we consider the achievable rate region of the Gaussian Multiple Access Channel (MAC) when suboptimal transmission schemes are employed. Focusing on the two-user MAC and assuming uncoded Pulse Amplitude Modulation (PAM), we derive a rate region that is a pentagon, and propose a strategy with which it can be achieved. We also compare the region with outer bounds and with orthogonal transmission.  
\end{abstract}
\IEEEpeerreviewmaketitle

\section{Introduction}

One of the scenarios to which the Multiple Access Channel (MAC) model applies is the uplink of cellular systems where multiple independent users/transmitters send information to a single base station/receiver. Therefore, the MAC is of significant importance in wireless communications. The capacity region of the Gaussian MAC with continuous input alphabets (Gaussian codebooks) is well known \cite{Cover06}. Although it provides insights into the achievable rate pairs in an information theoretic sense, it is based on the assumption of long Gaussian codebooks. In practice, there are constraints on the input alphabets and on the size of the codebooks. Practical communication systems employ signal constellations such as PAM, QAM and PSK and our focus in this paper is the simplest case of 2 users employing uncoded PAM transmission over the Gaussian MAC. 

The capacity region of the Gaussian MAC is a pentagon \cite{Cover06}, and can be achieved by single-user encoding and decoding combined with successive interference cancellation (SIC). Here superposition coding simplifies to each user transmitting a Gaussian codeword independently. The received signal is a sum of Gaussian codewords; the receiver performs successive decoding with interference cancellation or joint decoding. MAC was studied in \cite{Rajan08} for QAM alphabets showing the maximum achievable rate for these constellations. The achievable rate region was again shown to be a pentagon but the scope of this work was restricted to what best can be achieved by QAM alphabets in MAC (regardless of the codebook size) without elaborating how it can be achieved. We focus in this paper on the simplest practical scenario of uncoded PAM alphabets and look not only at the achievable rate region but also consider how this region can be achieved in a real world communication system. We have shown in this paper that because of the discrete alphabet constraint, in addition to superposition, power control is required to attain a rate region that has the form of a pentagon. Else, if power control is not used, the region reduces to a quadrilateral.

Focusing on the 2-user Gaussian MAC with uncoded PAM in this paper, our contributions are as follows: 
\begin{itemize}
\item We first derive inner and outer bounds for the achievable rate region of the 2-user Gaussian MAC when uncoded PAM is employed.
\item We show how the users can transmit non-integer rates when superposition is employed by also using power control. This results in an achievable rate region that is a pentagon.
\item We compare the achievable rate region with orthogonal transmission such as time division multiple access (TDMA) with power control and with the outer bounds and we conclude that the typical loss with respect to the outer bound (a Jensen loss) is small.

\end{itemize}

The paper is organized as follows. In Section \ref{sec:gap}, we review the rate penalty when discrete alphabets are employed for transmission over the single-user AWGN channel. Section \ref{sec:region} is devoted to the achievable rate region (inner and outer bounds) when uncoded PAM is used over the Gaussian MAC and power control is not applied, while Section \ref{sec:general} introduces power control to obtain a rate region in the form of a pentagon. Some examples of achievable rate regions are computed in Section \ref{sec:simulations}.

\section{Effect of discrete alphabets on the achievable rate over the single-user AWGN channel}

\label{sec:gap}

We first review the gap approximation for suboptimal transmission over the single-user AWGN channel. If the user employs PAM and has average power $P$, the minimum distance of the constellation is given by
\begin{align}
d_{\min}=\sqrt{12P/(M^{2}-1)} \label{eq:dmin_PAM}
\end{align}
If all the constellation points are equiprobable, the probability of symbol error is given by
\begin{align}
P_{e}=2\left(1-\frac{1}{M}\right)Q\left(\sqrt{\frac{3}{M^{2}-1}\mbox{SNR}}\right),\label{eq:Pe_PAM}
\end{align}
where $Q(x)=\int_{x}^{\infty}\frac{1}{\sqrt{2\pi}}e^{-\frac{u^{2}}{2}}du$ is the Q function and $\mbox{SNR}=\frac{P}{N_{0}}$ is the signal-to-noise ratio at the receiver. For a given $P_{e}$, by rearranging (\ref{eq:Pe_PAM}) we get
\begin{align}
\frac{3}{M^{2}-1}\mbox{SNR}&=\left[Q^{-1}\left(\frac{MP_{e}}{2(M-1)}\right)\right]^{2}\nonumber\\
\log_{2}M &=\frac{1}{2}\log_{2}\left(1+\frac{\mbox{SNR}}{\Gamma\left(\log_{2} M,P_{e}\right)}\right),
\end{align}
where $\Gamma\left(\log_{2} M,P_{e}\right)=\left[Q^{-1}\left(\frac{MP_{e}}{2(M-1)}\right)\right]^{2}/3$ is the gap (for PAM) \cite{Cioffi11}. Note that $\log_{2} M=R,$ where $R$ is the rate. 
The gap simplifies the calculation of achievable rates because it leads to an expression very similar to the elegant AWGN capacity formula. Computation is further simplified for the case of high SNR when the gap converges to $\Gamma_{\infty}\left(P_{e}\right)=\left[Q^{-1}\left(\frac{P_{e}}{2}\right)\right]^{2}/3$ as $M$ increases. The gap shrinks by the coding gain $\gamma$ once coded transmission is used, i.e. the new gap is given by $\Gamma/\gamma$, where $\gamma\leq\Gamma$.

\begin{figure}
\begin{center}
\includegraphics[scale=0.42]{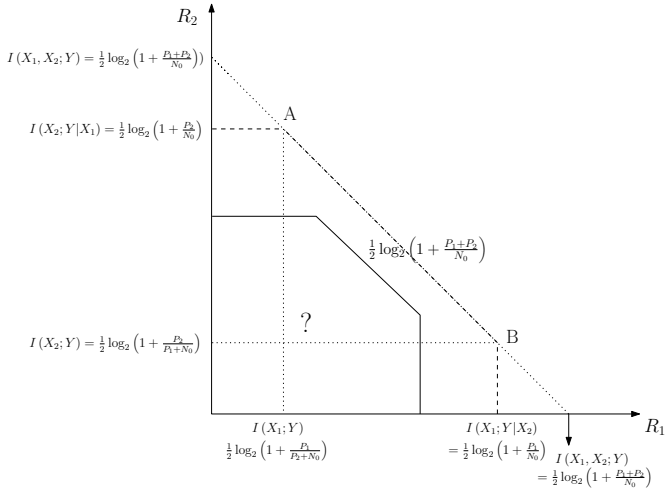}
\caption{\label{fig:MAC_region_PAM_Gaussian} The dashed line indicates the capacity region for Gaussian inputs; the continuous line indicates the rate region with discrete inputs but how this region can be achieved?}
\end{center}
\end{figure}

\section{Inner and outer bounds for the achievable rate region of the two-user Gaussian MAC with uncoded transmission}

\label{sec:region}

As is well known, the capacity region of the real-valued 2-user Gaussian MAC, illustrated in Fig.~\ref{fig:MAC_region_PAM_Gaussian} with dashed lines, is given by the following inequalities \cite{Cover06}:

\begin{align}
R_{1}&\leq\frac{1}{2}\log\left(1+\frac{P_{1}}{N_{0}}\right)=\frac{1}{2}\log\left(1+\mbox{SNR}_{1}\right),\nonumber\\
R_{2}&\leq\frac{1}{2}\log\left(1+\frac{P_{2}}{N_{0}}\right)=\frac{1}{2}\log\left(1+\mbox{SNR}_{2}\right), \ \mbox{and}\nonumber\\
R_{1}\!+\!R_{2}\!&\leq\!\frac{1}{2}\log\left(1\!+\!\frac{P_{1}+P_{2}}{N_{0}}\right)=\frac{1}{2}\log\left(1\!+\!\mbox{SNR}_{1}\!+\!\mbox{SNR}_{2}\right),\nonumber
\end{align}
where $\mbox{SNR}_{i}=P_{i}/N_{0}$.
The end points of this dashed line segment are achievable by single-user decoding combined with SIC, or by joint decoding \cite{ElGamal10}. Any point on the segment $AB$ other than the corners can be achieved by time sharing or rate splitting \cite{Urbanke96}.  

We now assume that uncoded PAM is employed by the users transmitting over the Gaussian MAC. Moreover, without loss of generality, let user 1 be stronger than user 2, i.e. $P_{1}\geq P_{2}$. The signal at the receiver is given by
\begin{align}
Y=X_{1}+X_{2}+Z,
\end{align}
where $X_{i}\in\mathcal{X}_{1}$ is the symbol sent by user $i$, and $M_{i}=\left|\mathcal{X}_i\right|$ is the cardinality of the PAM constellation of user $i$. $Z$ is Gaussian noise of variance $N_{0}$. 

Using the gap approximation, the achievable rate region can be upper-bounded by
\begin{align}
R_{1}&\leq \frac{1}{2}\log_{2}\left(1+\frac{P_{1}}{\Gamma\left(R_{1},P_{e}\right)N_{0}}\right), \label{eq:rate_user1}\\
R_{2}&\leq \frac{1}{2}\log_{2}\left(1+\frac{P_{2}}{\Gamma\left(R_{2},P_{e}\right)N_{0}}\right),\label{eq:rate_user2}\quad\mbox{and}\\
R_{1}+R_{2}&\leq \frac{1}{2}\log_{2}\left(1+\frac{P_{1}+P_{2}}{\Gamma\left(R_1+R_2,P_{e}\right)N_{0}}\right).\label{eq:rate_sum}
\end{align}
To simplify the discussion in the following, it is assumed that the values of the available powers $P_i$ are exactly the ones needed to achieve integer bit rates in (\ref{eq:rate_user1}) and (\ref{eq:rate_user2}), i.e. $P_i = \Gamma(k,P_e)\left(2^{2k}-1\right)N_0$ for some integer $k$. Then the rates given by the first two inequalities are achievable if only one user is transmitting. Moreover, they cannot be exceeded, since that would mean exceeding the maximum rate of the individual AWGN channels. 

For the third inequality, it is assumed that both users can cooperate. In this case, $P_1+P_2$ exceeds the power needed to transmit an integer number of bits. The bound in (\ref{eq:rate_sum}) cannot be exceeded, because that would mean that a user with power $P_{1}+P_{2}$ would be able to transmit above the maximum rate of the AWGN channel.

\begin{figure}
\begin{center}
\includegraphics[width=9cm]{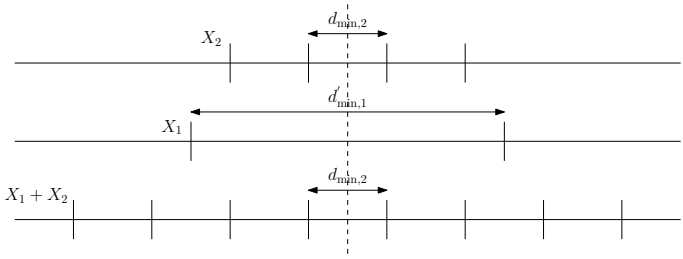}
\caption{\label{fig:sum_constellation_MAC} Superposition of two constellations at the receiver ensuring the required minimum distance.}
\end{center}
\end{figure}

\begin{figure}
\begin{center}
\includegraphics[scale=0.45]{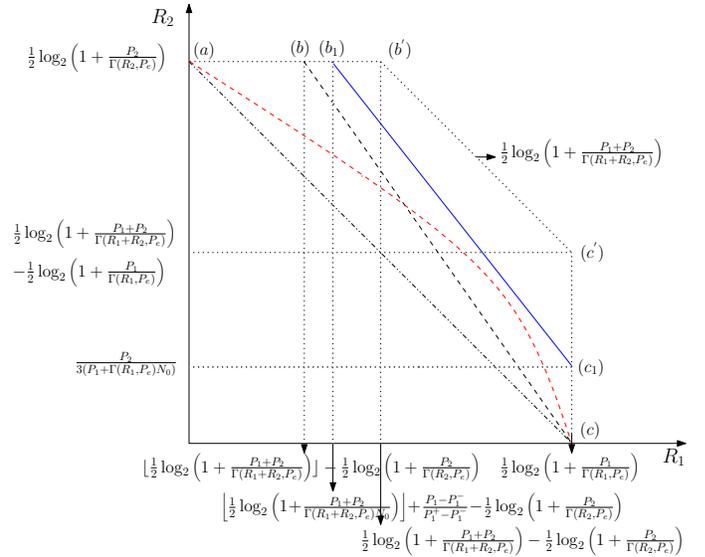}
\caption{\label{fig:MAC_region_PAM} Achievable rate regions for uncoded PAM and $P_i$ leading to integer bit rates. Outer bound (dotted line); superposition with power control (continuous blue line); TDMA with power control (dashed red curve); superposition without power control (dashed black line); naive TDMA, i.e. without power control (dashed-dotted line).}


\end{center}
\end{figure}

Note that, for sufficiently high SNR, one can write $P_i \approx 4^{R_{i}} \Gamma(R_{i},P_e) N_0$. Since it has been assumed that $P_1 \geq P_2$, it leads to
\begin{align}
 4^{R_1} \Gamma(R_{1},P_e) N_0 &< P_1 + P_2 < 4^{R_1+1} \Gamma(R_1,P_e) N_0  \nonumber\\ 
 &\leq 4^{R_1+1} \Gamma(R_1+1,P_e) N_0, \label{eq:rate_fullpower}
\end{align}
where we have used the fact that $\Gamma(R,P_e)$ is a nondecreasing function of the rate, $R$. We now show that 
\begin{align}
&R_{1}+R_{2}=\left\lfloor \frac{1}{2}\log_{2}\left(1+\frac{P_{1}+P_{2}}{\Gamma\left(R_1+R_2,P_{e}\right)N_{0}}\right)\right\rfloor \label{eq:sum_inner1}
\end{align}
is achievable. Obviously this is true when user 1 transmits at maximum rate as was shown in (\ref {eq:rate_fullpower}). Point $(c)$ of Fig.~\ref{fig:MAC_region_PAM} refers to this case. Assume now that user 2 transmits with full power and at maximum rate given by (\ref{eq:rate_user2}) (with equality), and forms a constellation of size $M_{2}=2^{R_2}$. Let the minimum distance between its constellation points be $d_{\min,2}$, which ensures the desired probability of error, $P_e$. From (\ref{eq:rate_user2}), the power of user 2 can be expressed as
\begin{align}
P_{2}=\Gamma\left(R_{2}, P_{e}\right)\left[2^{2R_2}-1\right]N_{0}.\label{eq:power}
\end{align}
The addition of $X_{1}$ at the receiver shifts $X_{2}+Z$ by a value that belongs to the constellation of user 1. Therefore, assuming $Z=0$, $X_{1}+X_{2}$ belongs to one of $M_{2}$ cosets, depending on the value of $X_{2}$. To ensure that the minimum distance $d_{\min}$ of the sum constellation ($X_{1}+X_{2}$) does not shrink beyond $d_{\min,2}$, user 1 needs to transmit with a minimum distance that is at least $M_{2}$ times the minimum distance of $X_{2}$, i.e. 
\begin{align}
d_{\min,1}\geq M_{2}d_{\min,2}.
\end{align}
This is illustrated in Fig.~\ref{fig:sum_constellation_MAC}. 
The total power, $P$, that is required to achieve the sum rate as given by (\ref{eq:sum_inner1}) satisfies $P \leq P_1 + P_2$. 
Therefore,
\begin{align}
P_1+P_{2}&\geq\Gamma\left(R_1+R_2,P_{e}\right)\left[2^{2(R_1+R_2)}-1\right]N_{0} \Rightarrow\nonumber\\
P_1 &\geq \Gamma\left(R_1+R_2,P_{e}\right)\left[2^{2(R_1+R_2)}-1\right]N_{0} \nonumber\\
&-\Gamma\left(R_{2},P_{e}\right)\left[2^{2R_2}-1\right]N_{0}\nonumber\\
&\geq \Gamma\left(R_{2},P_{e}\right)\left[2^{2(R_1+R_2)}-1\right]N_{0}\nonumber\\
&-\Gamma\left(R_{2},P_{e}\right)\left[2^{2R_2}-1\right]N_{0}\nonumber\\
		&\approx {2^{2R_2}}\Gamma\left(R_{1},P_{e}\right)\left[2^{2R_1}-1\right]N_{0},\label{eq:condition_P_1}
\end{align}
where we have assumed that the SNR is large enough so that $\Gamma\left(R_{2},P_{e}\right)\approx\Gamma\left(R_{1},P_{e}\right) \approx \Gamma_{\infty}(P_e)$, and have used the fact that, for fixed $P_{e}$, $\Gamma$ is a nondecreasing function of $R$. 
Therefore, $P_1$ is sufficient for user 1 to be able to transmit $M_1 = 2^{R_1}$ constellation points with minimum distance $M_{2}d_{\min,2}$ where $M_{2}=2^{R_{2}}$. Hence, point $(b)$ in Fig.~\ref{fig:MAC_region_PAM} is achievable. 


Unlike user 1, user 2 cannot achieve an integer number of bits. 
If user 1 transmits at its maximum rate as given by (\ref{eq:rate_user1}) (with equality), 
\begin{align}
P_{1}=\Gamma\left(R_{1},P_{e}\right)\left[2^{2R_1}-1\right]N_{0}.
\end{align}
Similar to point $(b)$, the achievability of the sum rate implies that
\begin{align}
P_{2} &\geq\Gamma\left(R_1+R_2,P_{e}\right)\left[2^{2(R_1+R_2)}-1\right]N_{0}-P_{1}\nonumber\\
      &=\Gamma\left(R_1+R_2,P_{e}\right)\left[2^{2(R_1+R_2)}-1\right]N_{0}\nonumber\\
			&\quad-\Gamma\left(R_{1},P_{e}\right)\left[2^{2 R_1}-1\right]N_{0}.
\end{align}
Because the minimum possible value of $R_{2}$ is 1, $\left[2^{2(R_1+R_2)}-1\right]\geq\left[4\cdot 2^{2R_1}-1\right]$. From the monotonicity of $\Gamma$ in $R$,
\begin{align}
P_{2}&\geq\Gamma\left(R_{1},P_{e}\right)\left[4 \cdot 2^{2 R_1}-1\right]N_{0}-\Gamma\left(R_{1},P_{e}\right)\left[2^{2 R_1}-1\right]N_{0}\nonumber\\
&=3P_{1}.
\end{align}
However, this contradicts the assumption that $P_{1}\geq P_{2}$. Thus, the weaker user cannot transmit an integer number of bits when the stronger user transmits at full rate, and $(c)$ cannot be exceeded. On the other hand, by the achievability of point $(b)$, similar to the case of optimal codebooks, even when the weaker user transmits at full rate, the stronger user can also transmit some data provided that the sum rate of (\ref{eq:sum_inner1}) exceeds the maximum achievable rate of user 2. Note that, when the values of the $P_i$ are exactly the ones that are needed to achieve an integer bit rate, the achievable sum rate given by (\ref{eq:sum_inner1}) is equal to the maximum rate that is achievable by (the stronger) user 1 as was shown in (\ref{eq:rate_fullpower}). 

The achievable points $(b)$ and $(c)$ can be joined by a straight line segment $\overline{bc}$ that corresponds to time sharing between these two points, similar to the case of points $A$ and $B$ in Fig.~\ref{fig:MAC_region_PAM_Gaussian}. In fact, we can improve time sharing between $(b)$ and $(c)$ by using power control. Nevertheless, as will be seen in Section \ref{sec:general}, it is possible to achieve the improved points $(b_1)$ and $(c_1)$ of Fig.~\ref{fig:MAC_region_PAM}, and it is, therefore, better to use time sharing between these improved points.

In Fig.~\ref{fig:MAC_region_PAM} the region obtained by orthogonal transmission (TDMA) with power control is also shown. As can be seen, in some cases, orthogonal transmission with power control may outperform the region that is achieved by superposition. However, the comparison is unfair in the sense that, in this section, power control was not used. In the following section, power control is combined with superposition. It is shown that this leads to an improved achievable region that has the form of a pentagon and appears to (always) contain the rate region obtained by TDMA with power control.

\section{An improved rate region achieved using superposition and power control}
\label{sec:general} 

Orthogonalization with power control relies on the fact that, because each user only transmits for part of the total time, the energy savings can be used to transmit with larger instantaneous power. The same principle can be applied to the superposition approach.

We begin by considering the case where user 1 is transmitting at full rate. Because it has been assumed that $P_{1}=\Gamma\left(R_{1},P_{e}\right)\left[2^{2R_1}-1\right]N_{0}$, the bound of (\ref{eq:rate_user1}) is achieved with equality. As shown in Section \ref{sec:region}, if user 2 were to transmit during the entire time, he would not be able to transmit any data (point $(c)$ of Fig.~\ref{fig:MAC_region_PAM}). However, by remaining silent for a fraction $\lambda_2$ of the total time, user 2 will then have accumulated enough energy to be able to superimpose a 2-PAM constellation on top of the constellation of user 1. Then for a fraction $1-\lambda_2$ of the total time, while the stronger user 1 still transmits at full rate, the weaker user 2 transmits 1 bit. 

One question that arises is whether user 2 should transmit a 2-PAM constellation or if it should accumulate even more energy and send a larger constellation for a smaller fraction of time. Because the rate of user 2 is equal to $\frac{1-\lambda_2}{2} \log_2\left(1+\frac{P_2}{(1-\lambda_2) \Gamma(R_2,P_e) N_0}\right)$, the largest possible $1-\lambda_2$ should be used, i.e. user 2 should transmit for more time 2-PAM instead of transmitting higher order modulation for less time.

$\lambda_2$ can be calculated easily by noting that, in order for user 2 to transmit 1 bit, the required power is equal to $d^2_{\min,2}/4$ (see (\ref{eq:dmin_PAM})). In order to preserve $P_e$, $d_{\min,2}$ should be at least equal to $M_1 d_{\min,1}$, which leads to 
\begin{align}
\frac{P_2}{1-\lambda_2} \geq \frac{3 P_1 M^2_1}{M^2_1-1} \Rightarrow \lambda_2 = 1-\frac{2^{2 R_1}-1}{2^{2 R_1}} \frac{P_2}{3 P_1}. \label{eq:user2_powerc}
\end{align}

Therefore, an achievable point, $(c_1)$, shown in Fig.~\ref{fig:MAC_region_PAM}, is $(R_1,\frac{2^{2 R_1}-1}{2^{2 R_1}} \frac{P_2}{3 P_1})$, with $R_1$ given by (\ref{eq:rate_user1}). This point can also be written as $(R_1,\frac{P_{2}}{3(P_{1}+\Gamma(R_{1},P_{e})N_{0})})$. Note that both users employ their entire energy to achieve this point.


The same approach can be used to improve point $(b)$ of Fig.~\ref{fig:MAC_region_PAM}. This time, the weak user 2 transmits with full power $P_2$ during the entire time, and attains (\ref{eq:rate_user2}) with equality. However, as discussed in Section \ref{sec:region}, $P_1+P_2$ exceeds the power that is required in order to achieve an integer rate. Hence, for a fraction of time equal to $\lambda_1$, user 1 can employ only the amount of power $P^{-}_1$ that is needed to achieve (\ref{eq:sum_inner1}) with equality (without floor operation). Then during the remaining time, it can boost its instantaneous power to the value $P^+_1$ required to transmit an additional bit. $P^{-}_1$ and $P^+_1$ can be obtained from (\ref{eq:condition_P_1}). If $\Delta P_1 \triangleq P^+_1 - P^-_1$ and $dP_1 \triangleq P_1 - P^-_1$, then from the power constraint,
\begin{align}
\lambda_1 P_{1}^{-}+\left(1-\lambda_1\right)P_{1}^{+}=P_{1}^{-}+dP_{1},
\end{align}
which leads to
\begin{align}
\lambda_1=1-\frac{dP_{1}}{P_{1}^{+}-P_{1}^{-}} = 1-\frac{dP_{1}}{\Delta P_1}. \label{eq:gamma2}
\end{align}

Hence, an improved point, $(b_1)$ can be achieved, as shown in Fig.~\ref{fig:MAC_region_PAM} given by $(\bar{R}_1,R_2)$, where $R_2$ is given by (\ref{eq:rate_user2}) and
\begin{align}
\bar{R}_1 \!=\! \left\lfloor\frac{1}{2}\log_{2}\left(1\!+\!\frac{P_{1}+P_{2}}{\Gamma\left(R_1+R_2,P_{e}\right)N_{0}}\right)\right\rfloor \!+\! \frac{P_{1}-P_{1}^{-}}{P_{1}^{+}-P_{1}^{-}}\!-\! R_2
\end{align} 
Note that first term in this equation is the maximum sum rate when power control is not used and is equal to the maximum rate of the strongest user, as discussed in Section \ref{sec:region}. Similar to point $(c_1)$, both users employ their entire energy to achieve $(b_1)$. Clearly, any point on the line segment $\overline{b_1 c_1}$ is also achievable by time sharing between the two extreme points.

We conclude by comparing the achievable sum rate to the outer bound given by (\ref{eq:rate_sum}). For point $(c_1)$, during the time fraction $(1-\lambda_2)$ when both the strong and the weak user transmit, the sum rate is equal to 
\begin{align}
\frac{1-\lambda_2}{2}\log\left(1 + \frac{P_1 + \frac{P_2}{1-\lambda_2}}{\Gamma(k+1,P_e)N_0}\right),
\end{align}
where $k$ is equal to $R_1$ as given by (\ref{eq:rate_user1}). By Jensen's inequality, adding the sum rates during two modes of operation and assuming that the SNR is large enough so that $\Gamma(k,P_e) \approx \Gamma_{\infty}(P_e)$, 
\begin{align}
&\frac{\lambda_2}{2}\log\left(1+\frac{P_1}{\Gamma_{\infty}(P_e)N_0}\right) + \frac{1-\lambda_2}{2}\log\left(1 + \frac{P_1 + \frac{P_2}{1-\lambda_2}}{\Gamma_{\infty}(P_e)N_0}\right)  \nonumber \\
&\leq\frac{1}{2}\log\left(1+ \frac{P_1 + P_2}{\Gamma_{\infty}(P_e)N_0}\right).
\end{align}

Hence, $(c_1)$ is below $(c')$. This loss occurs because of the need to transmit an integer number of bits during each fraction of time. The same conclusion can be reached for $(b_1)$. 

Because $P_1 \geq P_2$, the largest gap between the inner bound (\ref{eq:sum_inner1}) that corresponds to superposition without power control and the outer bound (\ref{eq:rate_sum}) occurs when $P_1=P_2$. This can be seen by writing 
\begin{align}
&\log\left(1 + \frac{P_1 + P_2}{\Gamma_{\infty}(P_e)N_0}\right) \nonumber \\
&= \log\left(1 + \frac{P_1}{\Gamma_{\infty}(P_e)N_0}\right) + \log\left(1 + \frac{P_2}{P_1 + \Gamma_{\infty}(P_e)N_0}\right),\nonumber
\end{align}
and noting that the first term is (\ref{eq:sum_inner1}).
Hence, the loss in sum rate with respect to the outer bound cannot exceed $\frac{1}{2}$ bit (under the assumption made in this paper that the $P_i$ have exactly the values needed for an integer number of bits per user). For large SNR, from (\ref{eq:user2_powerc}), $1-\lambda_2 \approx \frac{1}{3}$. Therefore, when the users have equal powers, the loss with respect to the outer bound, for large SNR, when the time sharing and power control scheme described in this section is used, is approximately equal to $\frac{1}{6}$ bit.

Throughout the paper it was assumed that the $P_i$ are exactly equal to the powers required to achieve an integer number of bits. In the more general case, time-sharing with power control can be employed to avoid underusing the available power. This case was not considered in this paper in order to present the main ideas and conclusions in a simpler setting.

\section{Simulation Results}

\label{sec:simulations}

\begin{figure}
\begin{center}
\includegraphics[width=9cm,height=7cm]{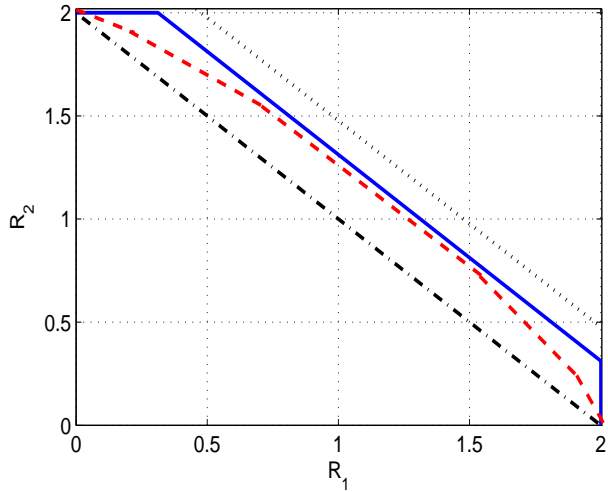}
\caption{\label{MAC_3e2_3e2_final} Outer bound and achievable rate regions for uncoded PAM with equal powers. $P_{1}=P_{2}=1.39\times 10^2$ and $N_{0}=1$. Outer bound (dotted line); superposition with power control (continuous line); TDMA with power control (dashed line); naive TDMA, i.e without power control (dashed-dotted line).}
\end{center}
\end{figure}

 \begin{figure}
\begin{center}
\includegraphics[width=9cm,height=7cm]{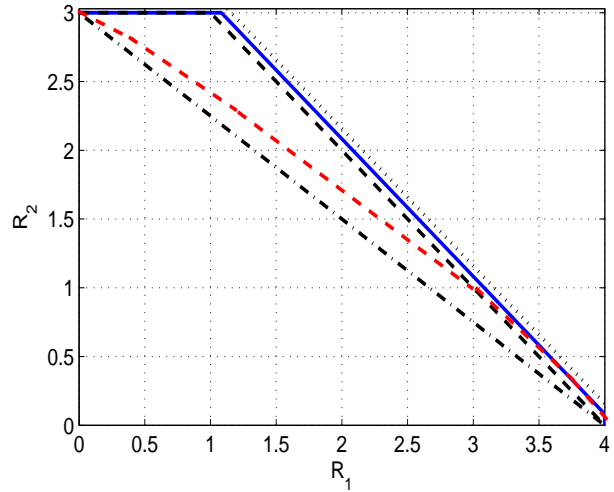}
\caption{\label{MAC_3e3_9e2} Outer bound and achievable rate regions for uncoded PAM with unequal powers. $P_{1}=2.4\times 10^3$, $P_{2}=5.9 \times 10^2$ and $N_{0}=1$. Outer bound (dotted line); superposition with power control (continuous line); superposition without power control (black dashed line); TDMA with power control (dashed red line); naive TDMA (dashed-dotted line).}
\end{center}
\end{figure}

We compute the rate regions that are achieved by different transmission schemes described in this paper. We look at different scenarios of power distribution between the two users. First we consider the case when both users have equal powers, which is shown in Fig.~\ref{MAC_3e2_3e2_final}. Superposition with power control achieves the largest achievable rate region, which is pentagon like the capacity region. Although TDMA with power control gets closer to the superposition region, it does not touch it due to the discrete nature of PAM alphabets. Note that, in the equal power case, superposition without power control collapses to the case of naive TDMA.  

We also consider the unequal power case shown in Fig.~\ref{MAC_3e3_9e2}, where user 1 is stronger than user 2. Here the achievable rate region with superposition coding employing power control almost touches the outer bound. Note that because user 2 is weaker, it can transmit a very small rate in the superposition mode. Superposition without power control collapses to a quadrilateral, as the weaker user cannot transmit when the stronger user transmits at full rate. TDMA with power control touches the boundary of superposition with power control, whereas naive TDMA is the most inferior among the transmission schemes. 

\bibliographystyle{IEEEtran}
\bibliography{temp}

\end{document}